# Copy Number Variants and Segmental Duplications Show Different Formation Signatures

Philip M. Kim[1], Jan Korbel[1], Xueying Chen[1] and Mark B. Gerstein[1,2]¶

[1]Department of Molecular Biophysics and Biochemistry
[2]Department of Computer Science
Yale University
New Haven, CT 06520

¶ To whom correspondence should be addressed.
Tel: +1 203 432 6105; Fax: +1 360 838 7861;
Email: mark.gerstein@yale.edu

## Abstract:

In addition to variation in terms of single nucleotide polymorphisms (SNPs), whole regions ranging from several kilobases up to a megabase in length differ in copy number among individuals. These differences are referred to as Copy Number Variants (CNVs) and extensive mapping of these is underway. Recent studies have highlighted their great prevalence in the human genome. Segmental Duplications (SDs) are long (>1kb) stretches of duplicated DNA with high sequence identity. They are generally thought to be the result of CNVs reaching fixation in the population.

To elucidate likely mechanisms of formation of these features, we examine in detail patterns of co-occurrence of different genomic features with both CNVs and SDs. First, we analyzed the co-localization of SDs and find that SDs are significantly co-localized with each other, resulting in a power-law distribution, which suggests a preferential attachment mechanism, i.e. existing SDs are likely to be involved in creating new ones nearby. This finding is further bolstered by an observation that young SDs most strongly co-localize with SDs that are closest in age and less strongly with older SDs. In line with this, we observe significant association of CNVs with SDs, but this association is weaker than may be expected.

Second, we look at the relationship of CNVs/SDs with various types of repeats. We we find that the previously recognized association of SDs with Alu elements is significantly stronger for older SDs and is sharply decreasing for younger ones. While it might be expected that the patterns should be similar for SDs and CNVs, we find, surprisingly, no association of CNVs with Alu elements. This trend is consistent with the decreasing correlation between Alu elements and younger SDs – the activity of Alu elements has been decreasing and by now it they seem no longer active. Furthermore, we find a striking association of SDs with processed pseudogenes suggesting that they may also have mediated SD formation. In line with the trend for Alu elements, this association is decreasing for recent SDs and is quite weak for CNVs. Moreover, find strong association with microsatellites for both SDs and CNVs that suggests a role for satellites in the formation of both. Finally, a manual analysis of a small number of CNV breakpoints is suggestive of an alternative mechanism.

In summary, we find striking differences in formation signatures of CNVs and SDs that can partly be explained by a decrease in Alu activity. We also find evidence for previously unrecognized mechanisms of CNV formation.

Keywords:

Segmental Duplication, Copy Number Variant, Non-allelic homologous recombination, Alu element, microsatellite, preferential attachment

## Introduction

With the rapid advances in high-throughput technology, the genomic study human genetic variation is emerging as a major research area. A large fraction of variation in terms of SNPs has been mapped and genotyped[1]. However, it has recently been recognized that a major fraction of mammalian genetic variation is manifested in an entirely different phenomenon known as copy number variation or structural variation. In contrast to SNPs or short indels, these variations correspond to large (1kb to several megabases) regions in the genome that are either deleted or amplified on certain chromosomes[2-6]. These Copy Number Variants (CNVs) have recently been mapped the 269 HapMap individuals. It is estimated to cover about 12% of the total human genome, thereby accounting for a major portion of human genetic variation[5]. Just like SNPs become fixed substitutions upon fixation, an (amplification) CNV that was fixed in the past is now visible in the genome as a Segmental Duplication (SD)[7]. A sizeable fraction (estimated to be 5.2%) of the human genome is covered in these SDs[7,8]. They are operationally defined as stretches of >1kb and 90% sequence identity and are especially widespread in the primate lineage[9]. Some annotated SDs may not be fixed in the population, but rather correspond to common CNVs, namely those that are present in the human reference genome. Hitherto, not much is known about mechanisms of CNV formation, but it has been suggested that non-allelic homologous recombination (NAHR) during meiosis can lead to the formation of larger deletions and duplications. NAHR is generally mediated by pre-existing repeats and Alu elements have been previously implicated in formation of SDs[10,11], whereas SDs have been suggested as mediating CNV formation[6,12]. A separate mechanism, non-homologous end joining (NHEJ) has been suggested for SD formation in subtelomeres[13].

In this work, we examine formation mechanisms of both SDs and CNVs in an integrated fashion. Specifically, we compare SDs and CNVs for co-localization with each other and repeat elements in the human genome. Our results show different signatures of formation for SDs and CNVs. While for SDs (and especially older ones), Alu mediated formation appears to be a predominant mechanism, we find little evidence for such a mechanism in CNVs. We present evidence for several alternative features that may contribute to the formation of both SDs and CNVs.

## Results

**Segmental Duplications follow a power law pattern in the human genome, suggesting a preferential attachment mechanism**

It has been suggested that CNV formation is partly mediated by NAHR mediated by Segmental Duplications[6,12,14]. If CNVs are indeed "SDs in the making", then these CNVs would later on, after fixation, be visible as SDs. We would then expect SD formation preferentially in regions with many previously existing SDs. This phenomenon represents one form of a preferential attachment mechanism ("rich get richer"). It is known that preferential attachment mechanisms generally lead to a power-law distribution[15]. Intuitively, a power-law or scale-free distribution corresponds to a distribution with a very long tail[16], i.e. there are a few regions in the genome which are extremely rich in Segmental Duplications. Hence, if SD-mediated NAHR would be a major factor contributing to new SDs, we would expect Segmental Duplications to be distributed according to a power-law throughout the human genome. Indeed, when analyzing different regions in the human genome for number of Segmental Duplications harbored, we observe a distinct power-law, which is suggestive of a preferential attachment mechanism (See Fig. 1). This result supports the theory that SD formation is indeed mediated by pre-existing SDs. The power-law distribution is independent of SD size, age, or the binning procedure (See Supplement).

**Segmental Duplications co-occur best with other Segmental Duplications of similar age**
Furthermore, an SD mediated NAHR mechanism would imply that recent SDs should co-occur with older Segmental Duplications. If we bin SDs according to sequence similarity (viewing sequence similarity as approximate age, since they diverge after duplication) we should see a significant co-occurrence between SDs most similar in sequence identity. Indeed, we observe a significant correlation between SDs of different sequence identity (See Fig. 2A and B). Strikingly, we observe that the best co-occurrence for any given age-bin is with the SDs in the next-older bin, consistent with a SD-mediated NAHR. Note that this result would also be consistent with different regions being susceptible to chromosomal rearrangements at different times. However, without a preferential attachment mechanism, it is very unlikely to observe a power-law distribution as in Fig 1. Finally, we observe that this correlation is best for old SDs and gets successively worse as we move towards more and more recent SDs. This may be indicative of a trend of changing SD formation behavior.

**The Alu mediated mechanism is complementary to the preferential attachment providing a complementary mechanism of SD generation**
As another mechanism for SD formation, NAHR mediated by Alu retrotransposons has been proposed[11]. We set out to examine this mechanism and find that indeed, SDs show highly significant co-localization with Alu elements (See Fig. 3A and Table S1), consistent with earlier reports[7,10]. Moreover, SDs appear to co-localize with L1 repeats, however, this association is much weaker and might just be reflective of co-localization of Alus and L1 repeats[17]. Therefore, as previously pointed out, Alu elements appear to be mediating SD formation. However, we also find that Alu mediated mechanisms and the preferential attachment mechanisms appear to be complementary. That is, SDs that co-localize strongly with Alus show weaker correlation with pre-existing SDs (See Fig 3B) than those that appear in Alu-poor regions. This result holds true for SDs of any sequence identity bin. This result suggests that a certain group of SDs is likely to have been formed by an Alu-mediated mechanism and another disjoint group is a more likely candidate for a mechanism involving pre-existing SDs.

**Processed pseudogenes show significant association with SDs and a small, but significant number of SDs are flanked by matching pseudogenes**
Processed pseudogenes were formed in a way similar to Alu retrotransposons, i.e., they utilize the same LINE retrotransposition machinery and are also thought to have been formed during the Alu burst ~40 Mya ago[18]. The obvious difference is that there is a much greater variety of pseudogenes than Alus elements. Therefore it is less likely that matching processed pseudogene pairs interact and to contribute to genome rearrangements by homologous recombination. However, we find a strong enrichment of pseudogenes with SDs. To evaluate whether these pseudogenes actually contributed to the formation of SDs, we performed a detailed breakpoint analysis of SDs. We find that in a small (144), but highly significant number of cases, we find matching processed pseudogenes at the matching SD junction regions of duplicated regions (See Fig 4). In an additional 78 cases, we find processed pseudogenes at both SD junctions that have different parent genes, but are highly similar (>95% sequence identity) over stretches of at least 200bp. This result presents evidence that pseudogenes did contribute to SD formation, albeit only in a small number of cases.

**Copy Number Variants co-occur with Segmental Duplications, but to a much smaller extent than might be expected**
It has been noted previously that Copy Number Variants also co-occur with Segmental Duplications and SD mediated NAHR has been suggested as a possible mechanism of CNV formation[6,12,19,20]. In light with this, CNVs have been viewed as the drifting, polymorphic form of SDs, i.e. SDs correspond to CNVs that have been fixed. This view, if true, would imply that CNVs should follow a similar pattern of distribution as very young SDs (i.e., SDs of very high sequence similarity), since they would have been created by similar mechanisms. We are, for simplicity, assuming a purely neutral model at this stage – but obviously SDs, as largely fixed genomic features may have undergone some kind of selection. In contrast, the largely polymorphic CNVs would only have to a much smaller extent, if only by virtue of being much younger. However, exploration of selection is beyond the scope of this study. Indeed, when analyzing SD and CNV distributions in the genome, we find, in line with previous results, that there is a significant overlap (See Fig. 5A and Table S1). However, what appears striking is that the correlation is not nearly as strong as the one observed for "young SDs" (>99% sequence identity), See Fig. 2B.

**Copy Number Variants do not show any significant association with Alu elements, but associate with microsatellite repeats and recombination hotpots**
If CNVs and SDs are formed by similar processes, one might assume that they also show association with Alu elements, as was the case for CNVs and pre-existing SDs. However, we find that CNVs show no significant association with Alu elements (See Fig. 5B). While there are weak associations with LINE repeats, these are due to the association of CNVs and SDs and disappear when we control for SD content (See Fig. 5C). This result implies that an Alu mediated mechanism is a very unlikely candidate for CNV formation. This result is consistent with reports that Alu mediated NAHR was most common during or shortly after the burst of Alu activity ~40Mya ago and has since declined[21]. Hence, the formation of CNVs and most SDs are probably mediated by different phenomena. One might argue that some of this difference is due to the different methods of experimental determination – SDs are read directly from the genome and CNVs used in this study are determined using microarrays. Therefore, we also computed associations between Alus and CNVs that were determined using two very different methodologies: Fosmid-paired end sequencing and genome assembly comparison (See Supplement). We find that a similar picture emerges: both techniques show essentially no significant association with Alu elements. Therefore, we conclude that Alu elements, while active in genome rearrangements in the past, do not currently play a role in formation of CNVs. It should be pointed out that this result does not contradict the notion of CNVs as drifting SDs – it simply means that the mechanism of CNV/SD formation has undergone significant change in the past 40 million years.

The absence of the association with Alu elements and the weakness of co-localization with SDs leads to the question of which genomic features are relevant for CNV formation. It has been suggested that microsatellite repeats have a role in mediation of chromosome rearrangements[22]. An association of SD junctions with microsatellites has previously been pointed out[11]. Hence, we examined whether they would associate with known CNVs. We indeed find that microsatellite repeats show a highly significant co-localization with CNVs (See Fig 5B and C), even after correcting for SD abundance. Finally, if homologous recombination events do lead to CNV formation one may expect an association with known recombination hotspots[1]. This analysis is confounded by the fact that duplication rich regions generally have a smaller coverage of SNPs and are hence less likely to show recombination hotspots due to this pre-ascertainment bias. In fact, we find that SD-rich regions are depleted in recombination hotspots (data not shown). However, when correcting for this bias, we find a weak, but significant association of CNVs with recombination hotspots (p-value of 0.001).

## Discussion

We have presented results that suggest changes in the formation of large genome rearrangements over the past 40 Mya. Our results suggest that shortly after the burst in Alu activity, an Alu or pseudogene mediated mechanisms were predominant in the formation of SDs. The formed SDs then presented highly homologous regions themselves and were active shortly after formation in generating new SDs. However, it is striking to see that the association of SDs with Alu elements is decreasing with decreasing age of the SD (increasing sequence similarity between the duplicates) (Fig 3C). Likewise, the co-localization of SDs with their younger counterparts is decreasing. These trends are indicative of a lesser contribution of homology mediated mechanisms for SD formation. At almost the same rate, preference of SDs for subtelomeric regions in the genome increases (Fig 3C). As was noted earlier, genesis of SDs in subtelomeric regions is largely due to a mechanism based on non-homologous end joining (NHEJ) mediated by micro-homologies (<25bp homology), rather than a NAHR mechanisms mediated by larger matching repeats[13].

For CNVs, a different picture emerges. The lack of association of CNVs with Alu elements is quite surprising, but is borne out by many different data sources. While it is possible that the low resolution of current CNV mapping techniques is partly responsible for this, the fact that SDs show a very strong association with the same coarse grained analysis appears to present evidence to the contrary. This lack of co-localization is in line with the emerging trend of decreasing Alu association of SDs. It is likely the result of continuing Alu divergence and hence, diminishing probability of Alu mediated NAHR. On the other hand it has previously been suggested that CNV associate with SD elements, and we find this trend to persist. However, SDs mediated CNV formation can only account for some of the CNVs found. Therefore, other mechanisms have to be at work as well. We suggest the following two possibilities for alternative mechanisms: First, the one repeat element that appears to associate strongly with CNVs are microsatellites. This co-localization might not be causal in nature, and only with higher resolution breakpoint data we will be able to answer this question with certainty. However, since microsatellites have been implicated in genome rearrangements, an involvement in CNV formation would certainly be sensible. Second, our findings are also suggestive of an increased role of NHEJ based mechanisms for the generation of CNVs which would also account for the poor association of CNVs with known repeats. Indeed, we find an association of CNVs towards subtelomeric regions (p-value<0.001), where double strand breakage and NHEJ is known to be prevalent. Obviously, more high-resolution information about CNV breakpoints is necessary to explore the mechanisms further; this data is expected to be available soon[23]. More breakpoint data could also shed light on the aforementioned role of microsatellites in CNV formation. As part of our pilot study, we manually curated 8 CNV breakpoints that have been identified and sequenced in the literature (See Table 1). The pattern that emerges is consistent with our proposed hypothesis: We see breakpoint microhomologies in 7 out of 8 breakpoints. On the other hand, only two of our 8 breakpoints overlap with Alu elements.

## Conclusions

We present evidence for different mechanisms for genesis of structural variants in the human genome. Our main result suggests that currently occurring Copy Number Variants appear to follow a decidedly different pattern than old Segmental Duplications. We show a clear shift from a prevalence of Alu-mediated generation of old SDs towards other mechanisms for more recent SDs. The surprising lack of association of CNVs with Alu elements can be viewed as the natural extension of this trend, as CNVs (that correspond to amplifications) are "very young" SDs. This trend is consistent with current models that propose a burst of Alu activity 40Mya ago and a subsequent decrease of Alu activity. We present results that suggest formation of CNVs through a number of alternative mechanisms, namely NAHR mediated by SDs and microsatellites as well as non-homologous end-joining at subtelomeres. Finally, we show a weak association of CNVs with recombination hotspots that might be indicative of another recombination based mechanism.

## Methods

### Sequence data preparation

We used the segmental duplications database from the University of Washington (http://humaparalogy.gs.washington.edu/dups) based on the build 35 genome[8]. We binned all existing SDs into sequence identity categories and different size categories (See Supplement). To enable comparison with low-resolution copy number variation data, we finally binned all segmental duplications according to genomic coordinate. We varied the binsize from 10kb to 1Mb. Because the copy number variant mapping resolution is at most 50kb for the techniques employed in the used datasets[24], we report the results for calculations with a binsize of 100kb. Calculations using other binsizes are reported in Table S1. For copy number variants we used three separate datasets, based on three different assay methodologies. The three-way comparison should avoid biases that may have been introduced by a single method. First, we used the recent set from the Human Copy Variation Consortium, which was based on microarray methods[5].

Secondly, the structural variation data based on Fosmid-paired-end sequencing was used[4]. Finally, a comparison of two different genome assemblies has revealed putative copy number variations[25]. The results from the latter two CNV datasets are reported in Table S1.

**Repeat Analysis**
Different kinds of repeats were identified using the genome annotation on the UCSC genome browser, based on the output of Repeatmasker. As above, distributions of Alu elements, LINE elements, and microsatellites were binned according to their genomic coordinates. Recombination hotspot data was taken from the hapmap recombination data[1]. Data for the processed pseudogenes was obtained from Pseudogene.org[26].

**Comparison of associations**
Coarse-grained co-localization was assessed by computing the spearman rank correlations between the binned distributions of each feature (SD occurrence, CNV occurrence or repeat occurrence) per bin. This measure is an accurate and robust measure of association and is independent of any assumptions of the distribution of the respective fatures. We used a binsize of 100kb for the analysis, but changes in the binning procedure did not have an effect on our results (See Supplement). This coarse grained approach can identify larger scale trends. It is especially suitable for the analysis of CNV associations because of the current low resolution mapping of their breakpoints. However, it may not be able to pinpoint exact breakpoint characteristics.

**Detailed SD breakpoint analysis for processed pseudogenes**
For a detailed analysis of processed pseudogene enrichment at SD breakpoints we analyzed all SD junctions for overlap with pseudogenes. Because of potential sequencing and alignment errors, we defined the SD junction as +/- 5 basepairs around the annotated breakpoint. We then looked for SDs where pseudogenes overlapped either the SD start or end junction in both duplicated segments. For each of these, we then compared the parent genes of the two pseudogenes that overlapped the SD junctions. For pseudogenes with different parent genes, we compared their sequence similarity using FASTA.

To assess the significance of the overlap between the processed pseudogenes and SD junctions, we first picked genomic regions of the same size and number as SDs at random and compared the overlap with processed pseudogenes. No matching junctions that had matching pseudogenes were found. As a second procedure that captures potential sequence biases, we randomized the SD junctions in a 5kb window around the actually junction and calculated the overlap with matching pseudogenes.

**Manual CNV breakpoint analysis**
To complement the coarse-grained approach, we manually analyzed a set of 8 manually curated CNV breakpoints. The CNV breakpoints were collected from the literature (all of them are based on PCR validated or cloned and resequenced breakpoints). The manual genomic analysis was carried out using the UCSC genome browser, the breakpoint neighborhood was examined for occurrence of repeated elements, recombination rates and other genomic features.

# Acknowledgments

We thank George Perry for careful reading of the manuscript and many insightful comments. We also thank Tara Gianoulis, Prianka Patel, Hugo Lam and Deyou Zheng for comments on the manuscript, technical assistance and helpful suggestions. This work was supported by the NIH.

# Tables

**Table 1:** Breakpoint regions of eight manually curated deletion CNV breakpoints from the recent literature[27-29]. We find microhomologies in 7 out of 8 breakpoints. Breakpoints #3 and #7 fall into Alu repeats at both ends, but show perfect homology only for 3 and 18 bp around the breakpoint. Breakpoint #6 shows no sign of homology. Microhomology regions are shown in red and the breakpoint is shown in bold. Chromosomal coordinates are given for hg17.

| ID | Source | Chr | Coordinates | Breakpoint region (left/right) |
|---|---|---|---|---|
| 1 | Newman et al. (2006) | 16 | 76929140 | GAGGATCTCAGTGCTCAGA A**T**GAA ATATCCCTGTACTCCAA |
|  |  |  | 76942399 | cttgtcattttattgggt a**t**gaa gcggtattgcattgtgat |
| 2 |  | 8 | 144771631 | TCCATCGAAAGGCGTTTA AA**A** GCAGTCACggccatgcgcgg |
|  |  |  | 144785838 | ACTTAGCAGCAAGGTCTC AA**A** CTTACATCCCAAAGTGACCT |
| 3 |  | 2 | 106338079 | gttcaagcgattctcccacc**tc**agcctcccaagtagctgga |
|  |  |  | 144785838 | cgagatcgcaccagtgcac**tc**cagcctgggcgagagagtga |
| 4 | Fan et al. (2005) | x | 146699050 | CCGCCTCTGAgcgggcggcg**gg**ccgacggcgagcgcgggcg |
|  |  |  | 146699260 | GAAGATGGAGGAGCTGGTG**T**GGAAGTGCGGGGCTCCAATG |
| 5 | Korbel et al. *in press* | 22 | 33969719 | ggattcaagtgattctcctg**cc**tcggcctcctgagtagctg |
|  |  |  | 33970693 | CCTACCCCACCACATTCCAA**CC**CACAGACAAGGACCAGCTC |
| 6 |  | 22 | 21548126 | GCGTGGGACAGCAGCACTGC**A**CACAGTGACACAGGCAGATG |
|  |  |  | 21566356 | AGCCTGTGTCACAGTGTGTG**G**TATTCGGCGGAGGGACCAAG |
| 7 |  | 22 | 17977963 | cactctgtcact caggctgg**a**gtgcagtggc acgatcccag |
|  |  |  | 19359814 | actgtgtcac caggctggag**t**gcagtggc tcaatcgtagct |
| 8 |  | 11 | 5203062 | ATCAAGCCTCTACTTGA ATC**C** TTTTCTGAGGGATGAATAAG |
|  |  |  | 5203681 | AATATGAAACCTCTTAC ATC**A** GTTACA atttatatgcagaa |

# Figure Captions

**Figure 1:** Segmental duplications are distributed according to a power-law in the human genome. As can be seen, segmental duplications follow a power-law distribution, i.e., while most regions in the genome are relatively poor in SDs, there is a small number of regions with much higher SD occurrence ($p(x) \sim x^{-0.31}$). This is indicative of a preferential attachment ("rich get richer") mechanism.

**Figure 2:** Heatmap of associations of SDs in different sequence identity bins. SDs co-occur best with pre-existing SDs of similar age and this trend appears be stronger for older SDs. Associations are given as spearman rank correlations of number of occurrence in genomic bins. All correlations are highly significant ($p\text{-value} \ll 0.00001$)

**Figure 3:** A) Alu mediated NAHR and preferential attachment are two complementary mechanisms for SD formation. In Alu rich regions (>10 Alu elements per 10kb), the association of SDs and pre-existing SDs is much lower than in Alu poor regions (No Alu elements per 100kb). Associations are given as spearman rank correlations of number of occurrence in genomic bins. All correlations are highly significant ($p\text{-value} \ll 0.00001$)
B) Association of Alu elements and SDs is highest for the oldest (~40Mya old) SDs and drops significantly for recent SDs. At the same time, preference for subtelomeric regions and a presumed NHEJ mechanism rises. Associations are given as spearman rank correlations of number of occurrence in genomic bins. All correlations are highly significant ($p\text{-value} \ll 0.00001$)

**Figure 4:** A) Pseudogene association with SDs. Just like Alu elements, pseudogenes co-localize very strongly with old SDs and less so with younger SDs. All correlations are highly significant ($p\text{-value} \ll 0.00001$)
B) Detailed SD junction analysis. A total of 144 SDs showed matching processed pseudogenes at both junctions, i.e. both peudogenes have the same parent gene and show high homology. When picking random genomic regions of the same size and number as SDs, no matching pseudogenes were ever found to overlap both SD junctions. When using an randomized offset of +/- 5kb to account for potential sequence biases, an average of 40 matching pseudogenes were found, but in 1000 trials, never more than 43.
C) Schematic of matching processed pseudogenes at SD junctions. The processed pseudogenes overlap matching SD junctions at both duplicated segments, making them likely candidates for having mediated NAHR.

**Figure 5:** A) Association of SDs and CNVs. CNVs co-localize with SDs, but much weaker than very young SDs. Associations are given as spearman rank correlations of number of occurrence in genomic bins. All correlations are highly significant ($p\text{-value} \ll 0.00001$) B) CNV association with different human repeat elements. CNVs associate weakly with L1 elements and microsatellites, but show no association with Alu elements. C) CNV association with human repeat elements after correcting for SD content. There is almost no significant association, the observed depletion in Alu elements may be due to a preference of CNVs for subtelomeric regions. Associations are given as spearman rank correlations of number of occurrence in genomic bins. p-values of the correlations are given in the bubbles.

# Figure 1

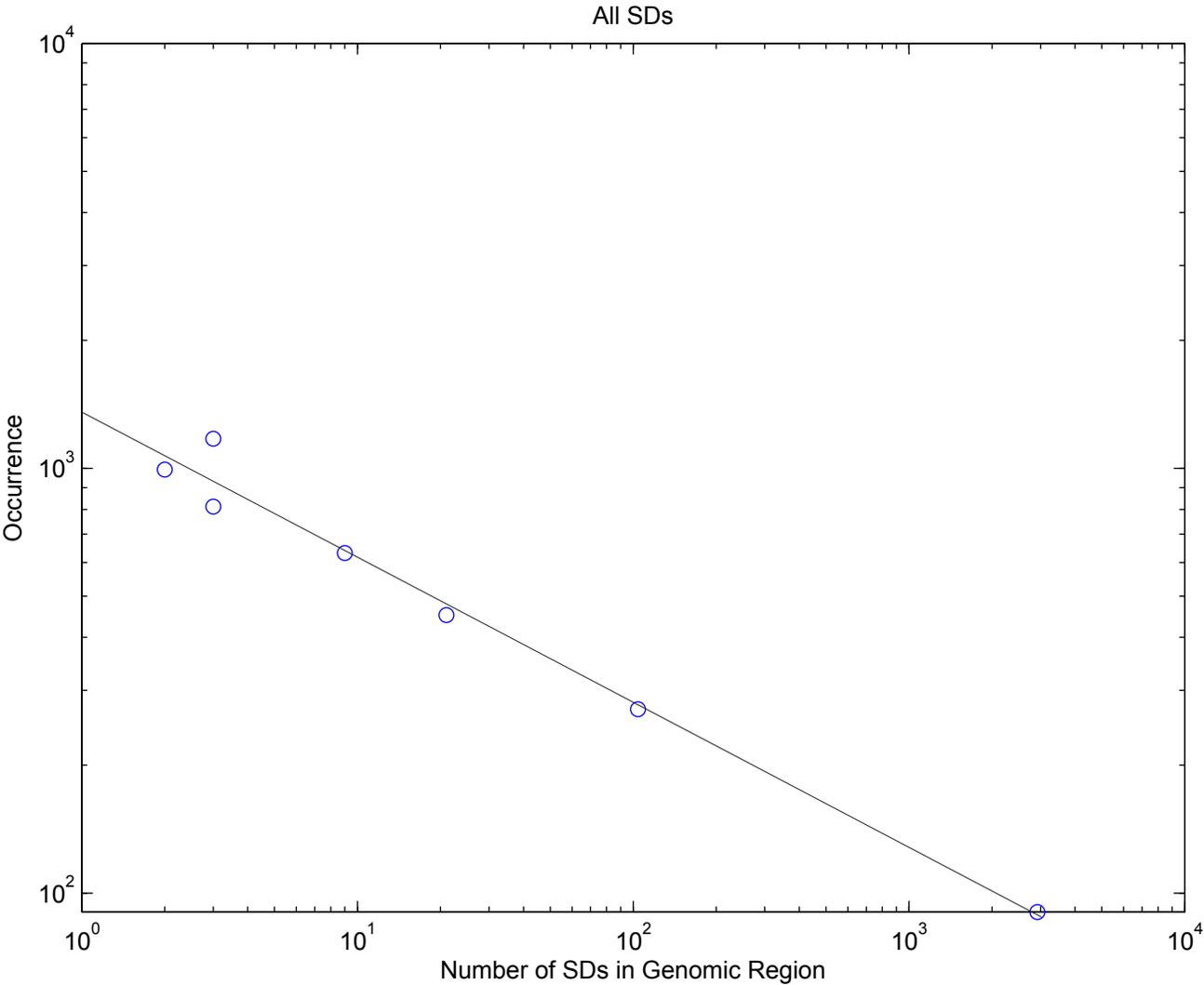

# Figure 2

## A

**SD/SD association with SDs by age**

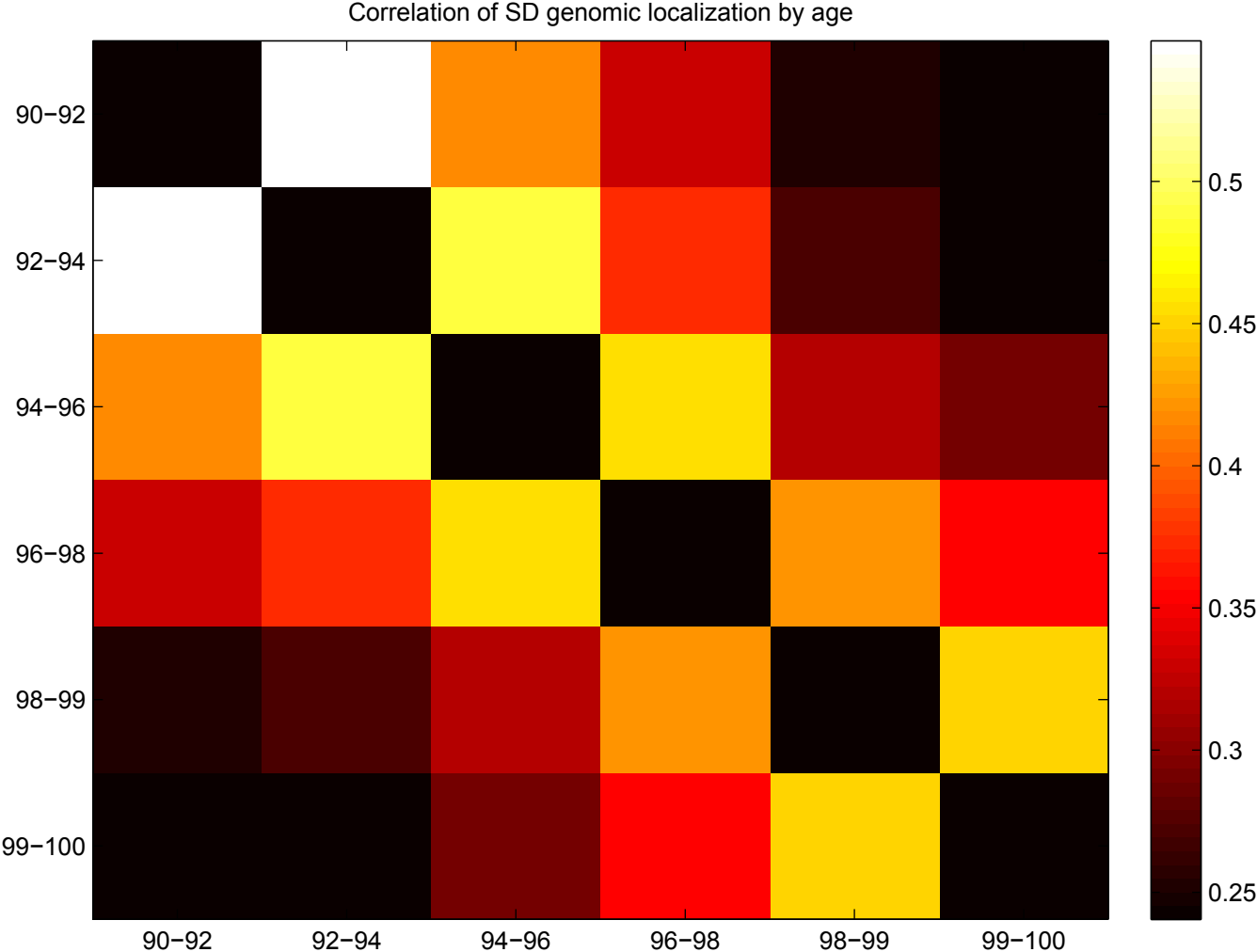

# Figure 3

## A

**SD (>99%) association with older SDs**

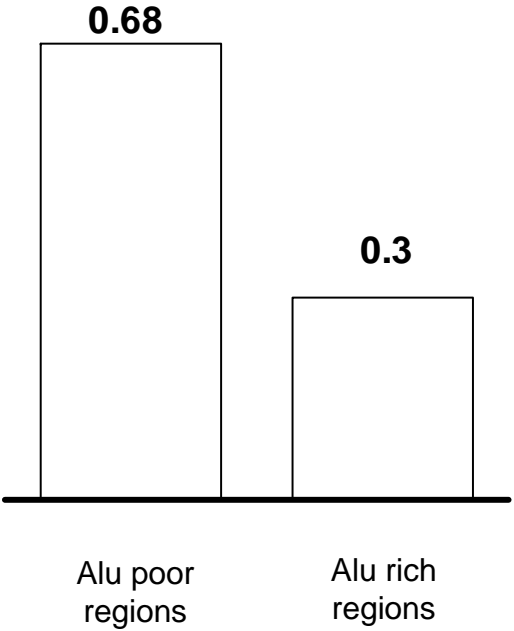

## B

**Alu association with SDs by age**

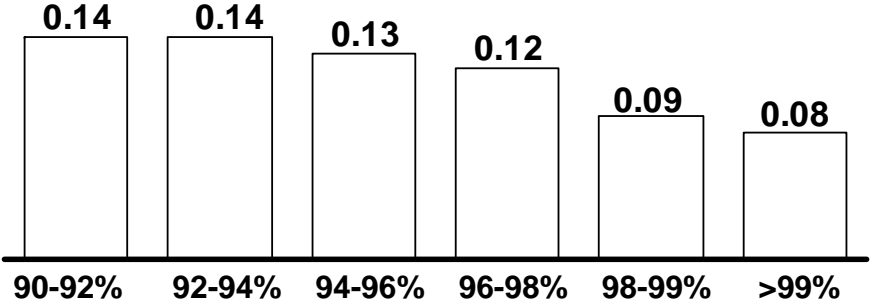

**SD association with subtelomeres**

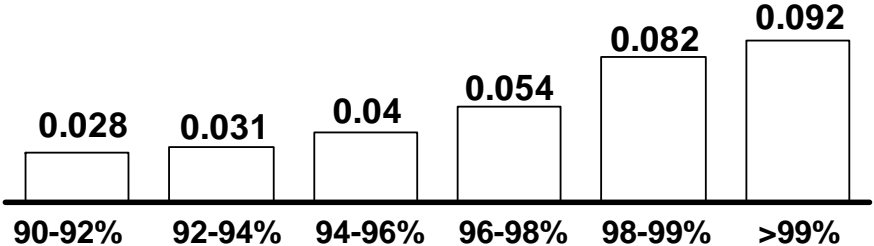

# Figure 4

## A

**Processed pseudogene association with SDs by age**

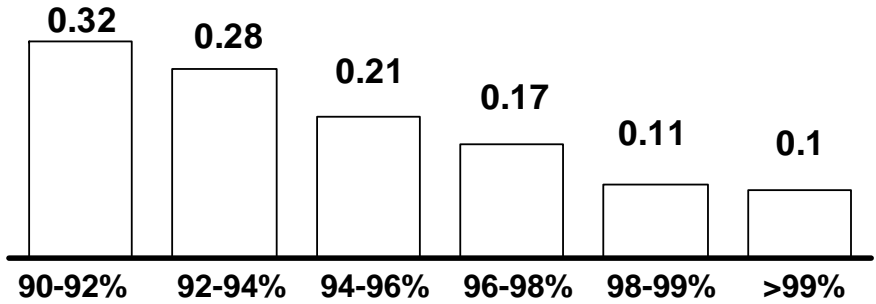

## B

**Processed pseudogenes at SD junctions**

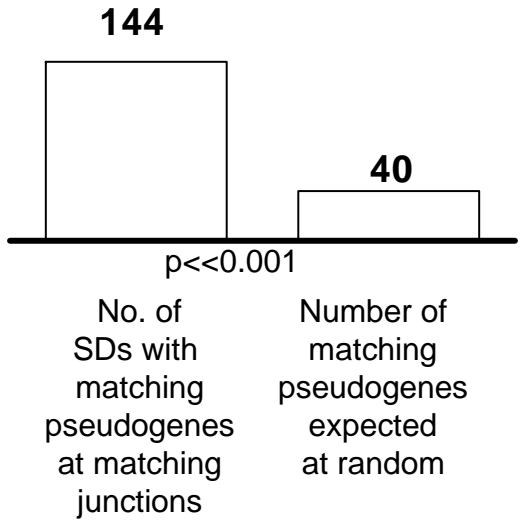

## C

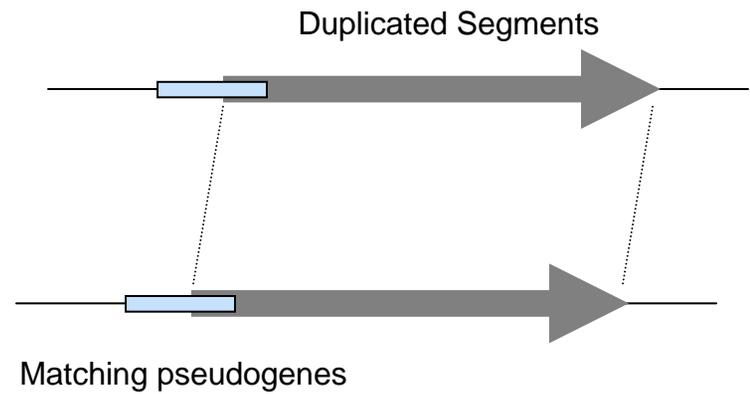

# Figure 5

## A
**Association of CNVs with SDs**

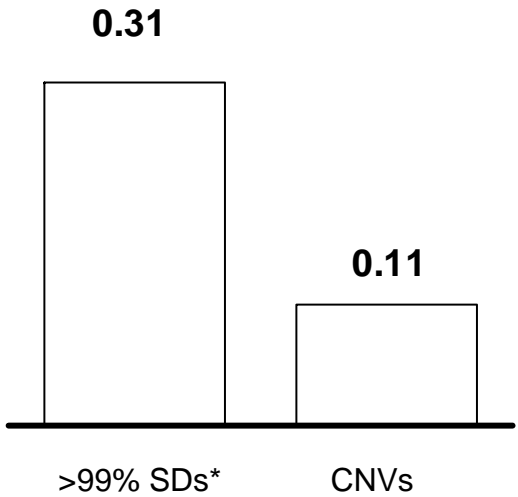

## B
**CNV association with repeats and processed pseudogenes**

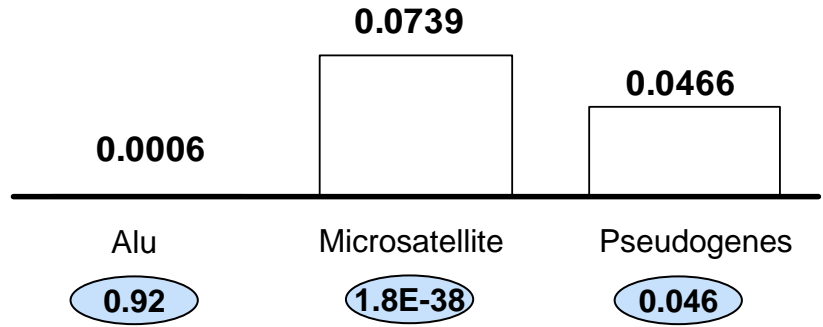

## C
**CNV association with repeats after correcting for SD content**

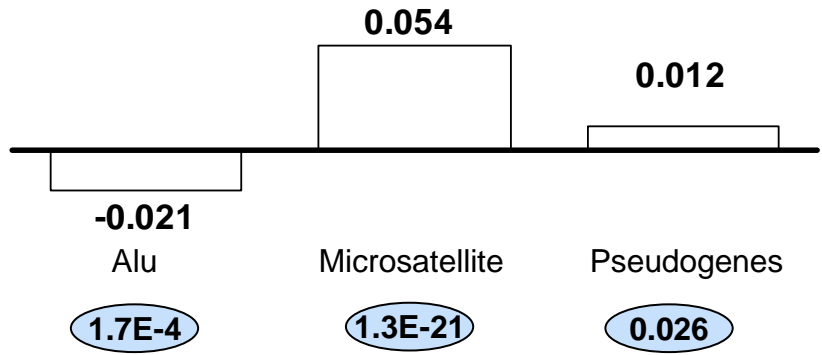